\documentclass{appolb}
\usepackage{epsfig}
\usepackage[T1]{fontenc}
\begin{document}
\title{Low energy neutrino scattering : from fundamental interaction
studies to astrophysics 
\thanks{Presented at the 45th Winter School in Theoretical Physics ``Neutrino Interactions: from Theory to Monte Carlo Simulations'', L\k{a}dek-Zdr\'oj, Poland, February 2--11, 2009.}
}
\author{Cristina Volpe
\address{Institut de Physique Nucl\'eaire, F-91406 Orsay cedex, 
CNRS/IN2P3 and University of Paris-XI, France}
}
\maketitle
\begin{abstract}
Neutrino scattering at low energies
is essential for a variety of timely applications potentially having 
fundamental implications, e.g.  
 unraveling unknown neutrino properties, such as the third
neutrino mixing angle, the detection of the diffuse supernova
neutrino background, or of cosmological neutrinos and furnishing a new constraint to
2$\beta$ decay calculations. Here we discuss some applications, the present status and the
perspectives.
\end{abstract}

\section{Introduction}
Core-collapse supernovae are massive stars that undergo gravitational
explosions at the end of their life, emitting most of their energy as
neutrinos of all flavours in a few seconds burst.
Neutrino scattering at low energy both on nucleons and on nuclei is important
for core-collapse supernova physics, not only for the observation of
the neutrinos emitted but also for processes occuring within the star
like  e.g. the nucleosynthesis of heavy elements (r-process) \cite{Balantekin:2003ip}. 
Both the explosion mechanism and the location of the r-process still need to be clarified. Measuring the neutrino luminosity curve produced during a future (extra-)galactic
explosion or of the diffuse supernova neutrino background (DSNB), from past explosions, can give essential
information both on the explosion, on unknown $\nu$ properties and on the star formation rate.

The present upper limits on the DSNB are furnished
by the Super-Kamiokande experiment, i.e. 1.2 $\bar{\nu}_{e}$
cm$^{-2}$s$^{-1}$ \cite{Malek:2002ns}
and 6.8 $\times$ 10$^3$ $\nu_e$
cm$^{-2}$s$^{-1}$  from LSD
\cite{Aglietta:1992yk},
at 90 $\%$ C.L.. Current theoretical predictions give strong indications that future observatories under study
\cite{Autiero:2007zj} (LAGUNA Design Study, in 2008-10 within FP7) should reach a sensitivity sufficient for a discovery
potential \cite{Ando:2004hc,Lunardini:2005jf,Horiuchi:2008jz}.
While (few) hundred events associated to inverse beta-decay are expected in water Cherenkov and
scintillator or to neutrino-argon scattering in argon detectors,  a few events on oxygen and carbon
can give
an improved limit compared to the LSD one, 
as first pointed out in \cite{Volpe:2007qx}. 

On the other hand,
one should not neglect alternative strategies to the
construction of large scale multipurpose (supernova neutrinos, proton decay and CP
violation) detectors. For example, as first suggested by Haxton and Johnson
\cite{Haxton:1987bf}, the measurement of $^{97}$Tc  produced by the 
$\nu$ scattering on $^{98}$Mo ore can allow the observation of galactic
neutrinos. Even though the idea is appealing, a recent
re-analysis considering our present knowledge on neutrino oscillations
in dense media has shown that  two unavoidable requirements are
an improved precision on the solar neutrino flux (a
significant background) and a precise knowledge of the
neutrino-nucleus cross sections \cite{Lazauskas:2009yh}. 

Serious improvements are currently made in the understanding of
core-collapse explosions on one hand and of neutrino
propagation in dense media on the other
\cite{Pantaleone:1992eq,Samuel:1993uw}. In particular,
a new paradigm has emerged due both to the inclusion of the
neutrino-neutrino interaction and of dynamical supernova density
profiles with shock waves. While the former engenders collective
phenomena \cite{Duan:2005cp,Hannestad:2006nj,Raffelt:2007xt}, the latter induces multiple resonances and phase
effects \cite{Schirato:2002tg,Fogli:2003dw,Dasgupta:2005wn,Kneller:2007kg}  (for a review see \cite{Duan:2009cd}).  
Recent works have explored possible direct
(in an observatory) or indirect (in the star) effects due to the possible
existence of CP violation in the lepton sector. They have established
that, indeed, there can be CP effects on the neutrino fluxes in a supernova
due to loop corrections or physics beyond the Standard Model \cite{Balantekin:2007es,Gava:2008rp}.  

In \cite{Engel:2002hg} 
a specificity of lead detectors is used : by using charged-current events
in conjunction with one- or two-neutron emission one can distinguish
between sin$^{2}\theta_{13}$>>10$^{-3}$ or  <<10$^{-3}$. While
the calculation performed needs further refinement, a future lead-based
detector -- HALO -- is now planned at SNOLAB. 
In \cite{Gava:2009pj} it has also
been shown that values of the third neutrino mixing
angle -- within or outside the experimental achievable range --
give a characteristic imprint  in the positron time signal
associated to inverse beta-decay. This calculation has the merit of
being the very first one putting together
the neutrino-neutrino interaction on one hand and the shock wave
effects on the other. In conclusion, the ever increasing level
of sophistication of supernova and neutrino propagation
modelling as well as of our knowledge of neutrino properties might
lead in the future to a concrete use of (relic) supernova neutrinos to unravel supernova physics
and/or unknown neutrino properties. 

A challenging application of (very) low energy neutrino capture on 
radioactive nuclei has been proposed very recently, the aim : 
detecting cosmological neutrinos \cite{Cocco:2007za}. Indeed, being non-relativistic 
at the present epoch, 
the cross sections,  with no reaction threshold, can strongly be enhanced. Extensive calculations over
thousands of nuclei have shown that one
might have a significant number of events per year. This idea
has been further investigated in \cite{Lazauskas:2007da,Blennow:2008fh}.

For the case of cosmological neutrinos the momentum transferred to the
nucleus is so low that experimental information from beta-decay
are used to avoid uncertainties inherent to nuclear structure calculations.
However, for neutrinos having energies in the 100 MeV energy range,
 the calculations present significant variations depending
on the details of the model and of the parametrization used... 

\section{Present status}
Computing neutrino-nucleus scattering cross sections in the several
tens MeV energy range requires modelling of the nuclear degrees of
freedom involving either isospin or spin-isospin transition matrix
elements. These calculations are sometimes particularly challenging
since they involve large model spaces and the inclusion of particular
configuration mixings or deformation. Another difficulty comes from the fact that in the
measurement: i) one cannot isolate the transition matrix elements
except in some specific cases (e.g. ground state to ground state
transitions); ii) one can only compare with convolved cross
sections. As a consequence, calculations that have significant discrepancies, at a given neutrino energy, 
can still be in a rather good agreement
with the measurements (if the achieved precision is not too good).
Note that so far only three nuclei have been studied experimentally, 
i.e. deuteron, carbon and iron. Many calculations exist based on
microscopic non-relativistic (such as e.g. \cite{Civitarese:2007ht}) or
relativistic approaches (see e.g. \cite{Paar:2008zza}).

The nuclear matrix elements, involved in the cross sections, are usually known as Fermi-type or Gamow-Teller type
transitions. The allowed transitions are rather well under control.
However a ``quenched'' axial-vector coupling constant is still used
to account for the difference between experimental and theoretical
matrix elements, calculated using various microscopic approaches, 
such as the Shell Model and the Quasi-particle Random-Phase
approximation. The forbidden transitions are still badly known.

The predictions tend to disagree as the neutrino impinging energy
increases. In some specific cases, such as the exclusive cross sections
on carbon, the discrepancies have been clarified \cite{Volpe:2000zn}, thanks also
to the wealth of experimental data available for this case. However
the inclusive cross sections are still not all understood. The
measurement performed on the iron nucleus, even though it
furnishes an important constraint, is not precise enough to discrimate
among various theoretical approaches \cite{Samana:2008pt}.
On the other hand related weak processes, like beta decay and muon
capture, or (Fermi and Ikeda) sum-rules help
keeping the theoretical ingredients under control. Still, we are far
from an accurate treatment of the  matrix elements
in the nuclei of interest such as $^{12}$C, $^{16}$O, $^{40}$Ar,
$^{56}$Fe, $^{98}$Mo, $^{97}$Tc and $^{208}$Pb.

Note that a better knowledge of the nuclear
response, and of forbidden transition, is particularly important 
for the double beta-decay searches  \cite{Volpe:2005iy}. In fact, one can
show that the nuclear matrix elements involved in the latter are the
same as those due to the exchange of a Majorana neutrino. Therefore
neutrino-nucleus experiments might furnish a supplementary constraint
to the half-life predictions that are still plagued by significant
variations. Obviously, since experiments cannot be
performed directly on the nuclei of interest, the 
calculations would benefit from an overall improvement of the nuclear
modelisation, e.g., from a step forward on the
quenching problem.

\section{Perspectives}
Experiments with future facilities can shed light on the weak nuclear
response in the several tens of MeV energy range. These are : 
low energy beta-beams \cite{Volpe:2003fi} -- that use the beta-decay of boosted radiocative ions \cite{Zucchelli:2002sa} --  
or facilities exploiting conventional sources (muon
decay-at-rest), such as $\nu$SNS \cite{nuSNS}, the European Spallation Source \cite{ESS}
facility, or at the future SPL proton driver. Neutrino-nucleus
interaction studies can be realized with detectors based on several
nuclei. Although the range of stable nuclei that can be investigated is
limited and the cross section measurements are inclusive, the
information obtained with such experiments would constrain the nuclear
ingredients of the microscopic approaches like the effective
interactions, the model spaces and the configuration mixings included.

The physics potential of low
energy neutrinos facilities covers a variety of 
aspects, from fundamental interactions to nuclear astrophysics
and nuclear structure studies. Low energy beta-beams might require either a devoted storage ring
\cite{Serreau:2004kx} or one/two detectors at off-axis
\cite{Lazauskas:2007va}. It has been shown that, at such a facility, 
a measurement of the
Weinberg angle at 10$\%$ precision level 
could improve  the LNSD measurement by about a
factor of 2 \cite{Balantekin:2005md}. A new test of the Conserved Vector Current
hypothesis and in particular of the weak magnetism contribution can also
be performed \cite{Balantekin:2006ga}. Other possibilities are  the study of non-standard interactions \cite{Barranco:2007ej}
and of coherent scattering \cite{Bueno:2006yq}. Besides, neutrino-nucleus measurements in the
100 MeV energy range can furnish information on the still badly known
forbidden transitions \cite{Volpe:2003fi,McLaughlin:2004va,Lazauskas:2007bs}. 
In Ref.\cite{Jachowicz:2006xx} the possibility of using a combination of beta-beam
spectra to analyse core-collapse supernova neutrino fluxes is proposed.
Note that most of
the ideas proposed for low energy beta-beams (for a review see \cite{Volpe:2006in}) 
are directly applicable
to spallation source facilities. 

In conclusion, pursuing low energy neutrino scattering studies, either
theoretical or experimental, with future spallation sources or low
energy beta-beams, can bring essential elements for our understanding
of the isospin and spin-isospin nuclear response, for core-collapse
supernova physics, for the weak interaction and neutrino physics.

\end{document}